\documentstyle[12pt]{article}

\def\ket#1{|#1\rangle}
\def\proj#1{|#1\rangle\langle#1|}
\def\outer#1#2{|#1\rangle\langle#2|}
\def\n{\vec n}
\def\alphaseq{\alpha_1,\ldots,\alpha_N}
\def\cseq{c_{\alpha_1\ldots\alpha_N}}
\def\nseq{(n_1)_{\alpha_1}\cdots (n_N)_{\alpha_N}}
\def\sigmaseq{\sigma_{\alpha_1}\otimes\cdots\otimes\sigma_{\alpha_N}}
\def\tr{{\rm tr}}
\def\Ket#1{|#1)}
\def\Bra#1{(#1|}
\def\Proj#1{|#1)(#1|}
\def\Inner#1#2{(#1|#2)}
\def\Outer#1#2{|#1)(#2|}
\def\G{{\cal G}}
\def\dOn{d\Omega_{\tilde n}\,}
\def\wc{w^{\rm can}}
\def\dOnAnB{d\Omega_{\n_A}d\Omega_{\n_B}\,}
\def\wnAnB{w(\n_A,\n_B)}
\def\PnAnB{P_{\n_A}\otimes P_{\n_B}}
\def\cat{\psi_{\rm cat}}
\def\rhoGHZ{\rho_{\epsilon{\rm GHZ}}}
\def\dO{d\Omega_{\n_A}d\Omega_{\n_B}d\Omega_{\n_C}\,}
\def\wn{w(\n_A,\n_B,\n_C)}
\def\Pn{P_{\n_A}\otimes P_{\n_B}\otimes P_{\n_C}}
\def\Pe#1#2#3#4#5#6{P_{#1\vec e_{#2}}\otimes
        P_{#3\vec e_{#4}}\otimes P_{#5\vec e_{#6}}}

\begin{document}

\title{Explicit product ensembles\\ for separable quantum states}
\author{
R\"udiger Schack\thanks{E-mail: r.schack@rhbnc.ac.uk} 
 $^{^{\hbox{\tiny (a,b)}}}$
and
Carlton M. Caves\thanks{E-mail: caves@tangelo.phys.unm.edu} 
 $^{^{\hbox{\tiny (a)}}}$ \\ $\;$ \\
$^{\hbox{\tiny (a)}}$Center for Advanced Studies, Department of Physics
and Astronomy, \\
University of New Mexico, Albuquerque, NM~87131--1156, USA \\
$^{\hbox{\tiny(b)}}$Department of Mathematics, Royal Holloway, \\ 
University of London, Egham, Surrey TW20 0EX, UK
}

\date{\today}
\maketitle

\begin{abstract}
We present a general method for constructing pure-product-state representations 
for density operators of $N$ quantum bits.  If such a representation has
nonnegative expansion coefficients, it provides an explicit separable ensemble
for the density operator.  We derive the condition for separability of a
mixture of the  Greenberger-Horne-Zeilinger state with the maximally mixed 
state.
\end{abstract}

\section{Introduction}

The state of a quantum system composed of $N$ subsystems is {\it separable\/}
if it can be written as a classical mixture or classical {\it ensemble\/} of 
tensor product states. Although it is straightforward to decide whether a 
pure state is separable, the same question is in general unsolved for mixed 
states.  A simple separability criterion is known only for two subsystems, 
and even then only if one of the subsystems is a two-level system (qubit) 
and the other one is at most a three-level system (qutrit) 
\cite{Peres1996a,Horodecki1996a} 

Recently it has been shown that there exists a neighborhood of the maximally
mixed state in which all states are separable \cite{Zyczkowski1998}. For $N$
qubits the size of this neighborhood decreases exponentially with the number
of qubits \cite{Braunstein1999a}. These results are interesting, among other
reasons, because NMR quantum computers \cite{Cory1997,Gershenfeld1997,Cory1998}
operate with states near the maximally mixed state.

In this paper we develop a general method for constructing pure-product-state 
representations for $N$-qubit density operators; when the expansion 
coefficients in such a representation are nonnegative, it provides an
explicit product ensemble for the density operator, and the state is 
separable.  In Sec.~\ref{sec:super}, we introduce a superoperator formalism 
that allows us to treat different discrete and continuous representations 
on the same footing. In Sec.~\ref{sec:pauli}, we apply the formalism to a 
discrete overcomplete operator basis, leading to a simple rederivation of 
the separability bound found in \cite{Braunstein1999a}.  
Section~\ref{sec:continuous} introduces a ``canonical'' continuous 
representation for an arbitrary density operator, and Sec.~\ref{sec:discrete} 
develops a method for deriving discrete representations from the continuous 
representation.   In Sec.~\ref{sec:spherical}, we show that the canonical
continuous representation is the unique representation containing only scalar 
and vector spherical-harmonic components; all other representations can be 
obtained by adding arbitrary higher-order spherical harmonics.  
Section~\ref{sec:epcat} applies the canonical continuous representation to
mixtures of the $N$-qubit ``cat state'' with the maximally mixed state.
In Sec.~\ref{sec:GHZ}, we derive the conditions for separability of a
mixture of the Greenberger-Horne-Zeilinger state \cite{Greenberger1990}
with the maximally mixed state.  Section~\ref{sec:discuss} concludes with
a brief discussion.

\section{Product-state representations and separable ensembles}
\label{sec:prodrep}

\subsection{Superoperator formalism}
\label{sec:super}

The set of linear operators acting on a $D$-dimensional Hilbert space ${\cal
H}$ is a $D^2$-dimensional complex vector space ${\cal L(H)}$. Let us introduce
operator ``kets'' $\Ket{A}=A$ and ``bras'' $\Bra{A}=A^\dagger$, distinguished
from vector kets and bras by the use of round brackets \cite{Caves1999a}. Then
the natural inner product on ${\cal L(H)}$, the trace-norm inner product, can
be written as $\Inner A B=\tr(A^\dagger B)$. The notation ${\cal S}=\Outer A B$
defines a superoperator ${\cal S}$ acting like
\begin{equation}
  \label{eq:super}
  {\cal S}\Ket X = \Ket A \Inner B X =\tr(B^\dagger X) A \;.
\end{equation}

Now let the set $\{\Ket{N_j}\}$ constitute a (complete or overcomplete) 
operator basis; i.e., let the operator kets $\Ket{N_j}$ span the vector 
space ${\cal L(H)}$. It follows that the superoperator $\G$ defined by
\begin{equation}
  \label{eq:g}
  \G \equiv \sum_j \Proj{N_j}
\end{equation}
is invertible.  The operators 
\begin{equation}
  Q_j\equiv\G^{-1}\Ket{N_j}
\end{equation}
form a dual basis, which gives rise to the following resolutions of the 
superoperator identity:
\begin{equation}
  \label{eq:resid}
  {\bf 1} = 
  \sum_j \Outer{Q_j}{N_j} =
  \sum_j \Outer{N_j}{Q_j}
\;.
\end{equation} 
An arbitrary operator $A$ can be expanded as
\begin{equation}
  \label{eq:expansion1}
  A = \sum_j \Outer{N_j}{Q_j} A) = 
      \sum_j N_j \tr(Q_j^\dagger A) 
\end{equation}
and
\begin{equation}
  \label{eq:expansion2}
  A = \sum_j \Outer{Q_j}{N_j} A) = 
      \sum_j Q_j \tr(N_j^\dagger A) \;.
\end{equation}
These expansions are unique if and only if the operators $N_j$ are linearly
independent.

\subsection{A discrete representation for the 6 cardinal directions}
\label{sec:pauli}

The Hermitian operators $P_{j\mu}$, $j=1,2,3$, $\mu=0,1$, defined by 
\begin{equation}
  \label{eq:pauli}
  P_{j\mu} \equiv {1\over2}\Bigl(1+(-1)^\mu\sigma_j\Bigr)\;,
\end{equation}
are the pure-state projectors corresponding to the 6 cardinal directions
on the Bloch sphere.  They form an overcomplete basis in the space of 
operators acting on a qubit.  The superoperator
\begin{equation}
  \G = \sum_{j,\mu} \Proj{P_{j\mu}} = 
  {1\over2} \Biggl( 3 \Proj1 + \sum_j \Proj{\sigma_j}\Biggr)
  \label{eq:sixG}
\end{equation}
has normalized eigenoperators $\Ket1/\sqrt2$ and $\Ket{\sigma_j}/\sqrt2$,
with eigenvalues 3 and 1, respectively, from which it follows that
\begin{equation}
  \G^{-1} = {1\over2}\Biggl( {1\over3}\Proj1 + \sum_j\Proj{\sigma_j}\Biggr)\;.
\end{equation}
The last equation allows us to find the dual-basis operators,
\begin{equation}
  \label{eq:paulidual}
  Q_{j\mu} = \G^{-1} \Ket{P_{j\mu}} 
  = {1\over6}\Bigl(1 + 3(-1)^\mu\sigma_j\Bigr) \;.
\end{equation}
Any density operator $\rho={1\over2}(1+\vec S\cdot\vec\sigma)$ for one qubit 
can be represented in the form
\begin{equation}
  \label{eq:rhopauli}
  \rho = \sum_{j,\mu} \Ket{P_{j\mu}}\Inner{Q_{j\mu}}\rho =
  {1\over6} \, \sum_{j,\mu} P_{j\mu}\Bigl(1+3(-1)^\mu S_j\Bigr)\;. 
\end{equation}

For an $N$-qubit density operator, the analogous representation is
\begin{equation}
  \label{eq:rhonpauli}
  \rho = 
  \sum_{j_1,\mu_1,\ldots,j_N,\mu_N} \wc(j_1,\mu_1,\ldots,j_N,\mu_N)
    P_{j_1\mu_1}\otimes\cdots\otimes P_{j_N\mu_N} \;,
\end{equation}
where
\begin{eqnarray} 
  \wc(j_1,\mu_1,\ldots,j_N,\mu_N) &\equiv&
    \tr(\rho\,Q_{j_1\mu_1}\otimes\cdots\otimes Q_{j_N\mu_N}) \nonumber \\
    &=& {1\over6^N}\tr\biggl(\rho
      \Bigl(1+3(-1)^{\mu_1}\sigma_{j_1}\Bigr)
      \otimes\cdots\otimes
      \Bigl(1+3(-1)^{\mu_N}\sigma_{j_N}\Bigr)\biggr)\;. \nonumber \\
  \label{eq:wnpauli} 
\end{eqnarray}
This representation exists for arbitrary density operators, entangled or
separable. If the coefficients $\wc(j_1,\mu_1,\ldots,j_N,\mu_N)$ are
nonnegative, the density operator $\rho$ is separable, and the representation 
(\ref{eq:rhonpauli}) provides an explicit product-state ensemble for the 
density operator.  The expansions~(\ref{eq:rhopauli}) and (\ref{eq:rhonpauli}) 
are not unique because the 6 operators $P_{j\mu}$ are not linearly independent.
Throughout the paper we use the superscript ``can,'' standing for 
{\it canonical}, to denote the natural, but generally not unique expansion 
coefficients that come from using the procedure of Sec.~\ref{sec:super}. 

For any $N$-qubit density operator $\rho$, the coefficients~(\ref{eq:wnpauli}) 
obey the bound
\begin{equation}
  \label{eq:bound1}
  \wc(j_1,\mu_1,\ldots,j_N,\mu_N) \ge 
  \pmatrix{\hbox{smallest eigenvalue of}\cr
           Q_{j_1\mu_1}\otimes\cdots\otimes Q_{j_N\mu_N}}
  = -{2^{2N-1}\over6^N} \;.
\end{equation}
This follows from the fact that $Q_{j\mu}$ has eigenvalues $2/3$ and $-1/3$. 
The most negative eigenvalue of the product operator
$Q_{j_1\mu_1}\otimes\cdots\otimes Q_{j_N\mu_N}$ is therefore 
$(-1/3)(2/3)^{N-1}=-2^{2N-1}/6^N$, from which Eq.~(\ref{eq:bound1}) follows. 

Consider now an $N$-qubit density operator that is a mixture of the maximally
mixed density operator, $1/2^N$, and an arbitrary density operator $\rho_1$:
\begin{equation}
  \rho_\epsilon = {1-\epsilon\over2^N}1 + \epsilon\rho_1 \;.
  \label{eq:rhoeps}
\end{equation}
Any density operator can be written in this form \cite{Braunstein1999a}. 
The coefficients~(\ref{eq:wnpauli}) for $\rho_\epsilon$ are given by
\begin{eqnarray}
  \wc_\epsilon(j_1,\mu_1,\ldots,j_N,\mu_N) &=& 
  {1-\epsilon\over6^N} + \epsilon \wc_1(j_1,\mu_1,\ldots,j_N,\mu_N) \nonumber\\
  &\ge& {1-\epsilon(1 + 2^{2N-1})\over6^N} \;,
\end{eqnarray}
which is nonnegative for
\begin{equation}
  \label{eq:bound2}
  \epsilon \le {1\over1+2^{2N-1}} \;.
\end{equation}
This bound, derived in \cite{Braunstein1999a} using a different route, 
implies that, for $\epsilon \le {1/(1+2^{2N-1})}$, all density operators 
of the form (\ref{eq:rhoeps}) are separable. It sets a lower bound on the 
size of the separable neighborhood surrounding the maximally mixed state.

\subsection{Canonical continuous representation}
\label{sec:continuous}

The set of all pure-state projectors on the Bloch sphere,
\begin{equation}
  P_{\n} \equiv |\n\rangle\langle\n| = {1\over2}(1+\vec\sigma\cdot\n) \;,
\end{equation}
forms an overcomplete basis. The corresponding superoperator
\begin{equation}
  \label{eq:contG}
  \G = \int d\Omega\,\Proj{P_{\n}} =
  \pi \Biggl( \Proj1 + {1\over3} \sum_j \Proj{\sigma_j}\Biggr) 
\end{equation}
is proportional to the superoperator of Eq.~(\ref{eq:sixG}).  The normalized
eigenoperators of $\G$ are $\Ket1/\sqrt2$ and $\Ket{\sigma_j}/\sqrt2$, 
with eigenvalues $2\pi$ and $2\pi/3$, respectively. Thus $\G^{-1}$ can be 
expressed as 
\begin{equation}
  \G^{-1} = {1\over4\pi}\!\left( \Proj1 + 3\sum_j\Proj{\sigma_j} \right) \;.
\end{equation}
The dual-basis operators are then given by 
\begin{equation}
  \label{eq:contdual}
  Q_{\n} = \G^{-1} \Ket{P_{\n}} 
  = {1\over4\pi} (1+3\vec\sigma\cdot\n) \;.
\end{equation}
Any density operator $\rho={1\over2}(1+\vec S\cdot\vec\sigma)$ for one qubit 
can be represented in the form
\begin{equation}
  \label{eq:rhocont}
  \rho = \int d\Omega\, \Ket{P_{\n}}\Inner{Q_{\n}}\rho =
      {1\over4\pi}\int d\Omega\,P_{\n}(1+3\vec S\cdot\n) \;.   
\end{equation}

For $N$ qubits, we define the pure-product-state projector
\begin{equation}
  \label{eq:prodproj}
  P(\tilde n)\equiv P_{\vec n_1}\otimes\cdots\otimes P_{\vec n_N}
  = {1\over2^N}(1+\vec\sigma\cdot\n_1)\otimes\cdots
  \otimes(1+\vec\sigma\cdot\n_N) \;,
\end{equation}
where $\tilde n$ stands for the collection of unit vectors 
$\vec n_1,\ldots,\vec n_N$.  Any $N$-qubit density operator can be expanded as
\begin{equation}
  \rho = \int \dOn \wc(\tilde n) P(\tilde n) 
  \equiv\int d\Omega_{n_1}\cdots d\Omega_{n_N}\wc(\tilde n) P(\tilde n) \;,
  \label{eq:rho}
\end{equation}
where
\begin{equation}
  \wc(\tilde n) \equiv \tr\Bigl(\rho Q(\tilde n)\Bigr)
  \label{eq:wn}
\end{equation}
is the canonical expansion function, with 
\begin{equation}
  Q(\tilde n) \equiv Q_{\vec n_1}\otimes\cdots\otimes Q_{\vec n_N} =
  {1\over(4\pi)^N} (1+3\vec\sigma\cdot\n_1)\otimes\cdots
  \otimes(1+3\vec\sigma\cdot\n_N) \;.
\end{equation}
If the coefficients $\wc(\tilde n)$ are nonnegative, the density operator 
$\rho$ is separable, and the representation (\ref{eq:rho}) provides an 
explicit product-state ensemble for the density operator.  

Using the same argument as in the preceding subsection, one can show that 
for any density operator $\rho$ \cite{Braunstein1999a},
\begin{equation}
  \label{eq:bound3}
  \wc(\tilde n) \ge 
  \pmatrix{\hbox{smallest eigenvalue}\cr
           \hbox{of $Q(\tilde n)$}}
  = -{2^{2N-1}\over(4\pi)^N} \;.
\end{equation}
Applying this result to density operators of the form~(\ref{eq:rhoeps}), 
one arrives again at the lower bound~(\ref{eq:bound2}) on the size of the 
separable neighborhood of the maximally mixed density operator 
\cite{Braunstein1999a}.

\subsection{Other discrete representations}
\label{sec:discrete}

The expansion coefficients~(\ref{eq:wnpauli}) for the discrete representation
of Sec.~\ref{sec:pauli} can be obtained by evaluating the expansion 
coefficients~(\ref{eq:wn}) for the canonical continuous representation along 
the  cardinal directions for each qubit and then renormalizing the resulting 
function.  This is no accident.  It is easy to characterize a class of 
discrete representations whose expansion coefficients are similarly related 
to the continuous representation.

Consider a discrete set of projection operators,
\begin{equation}
  P_{\n_\alpha}=|\n_\alpha\rangle\langle\n_\alpha|=
  {1\over2}(1+\vec\sigma\cdot\n_\alpha)\;,\;\;\;
  \alpha=1,\ldots,K.
\end{equation}
The corresponding superoperator
\begin{eqnarray}
  \G &=& \sum_{\alpha=1}^K
  \Proj{P_{\n_\alpha}} \nonumber \\
  &=& {1\over4}\!\left(
  K\Proj 1 +
  \sum_\alpha\Bigl(\n_\alpha\cdot\Outer{\vec\sigma}{1}
                   +\Outer{1}{\vec\sigma}\cdot\n_\alpha\Bigr)
  +\sum_{j,k}\Outer{\sigma_j}{\sigma_k}
             \sum_\alpha(n_\alpha)_j(n_\alpha)_k
  \right)
  \label{eq:discreteG}
\end{eqnarray}
generates dual-basis operators and expansion coefficients proportional
to those for the continuous representation if and only if $\G$ is 
proportional to the superoperator~(\ref{eq:contG}) for the continuous 
representation, i.e., if and only if 
\begin{eqnarray}
  0&=&\sum_\alpha\n_\alpha \;, \label{eq:cond1} \\ 
  {1\over3} \, \delta_{jk}&=&{1\over K}\sum_\alpha(n_\alpha)_j(n_\alpha)_k \;. 
  \label{eq:cond2}
\end{eqnarray}
The trace of Eq.~(\ref{eq:cond2}) is satisfied by any collection of unit 
vectors.

When these conditions are satisfied, the superoperator~(\ref{eq:discreteG})
simplifies to
\begin{equation}
  \G = {K\over4} 
  \Biggl( \Proj1 + {1\over3} \sum_j \Proj{\sigma_j}\Biggr) \;,
\end{equation}
with an inverse
\begin{equation}
  \G^{-1}= {1\over K} 
  \Biggl( \Proj1 + 3 \sum_j \Proj{\sigma_j}\Biggr) 
\end{equation}
that generates dual-basis operators
\begin{equation}
  Q_{\n_\alpha} = \G^{-1}\Ket{P_{\n_\alpha}}
  = {1\over K}(1 + 3\vec\sigma\cdot\n_\alpha) \;.
\end{equation}
Any density operator $\rho={1\over2}(1+\vec S\cdot\vec\sigma)$ for one qubit 
can be represented in the form
\begin{equation}
  \label{eq:rhodiscrete}
  \rho = \sum_\alpha \Ket{P_\alpha}\Inner{Q_\alpha}\rho =
  {1\over K} \, \sum_{\alpha} P_{\alpha}(1+3\vec S\cdot\n_\alpha)\;. 
\end{equation}

The conditions~(\ref{eq:cond1}) and (\ref{eq:cond2}) are satisfied by unit 
vectors that point to the vertices of any regular polyhedron inscribed within 
the Bloch sphere---i.e., a tetrahedron, octahedron, cube, icosahedron, or 
dodecahedron.  The 4 projectors for a tetrahedron are linearly independent, 
making the corresponding tetrahedral representation~(\ref{eq:rhodiscrete}) 
unique.  An octahedron gives rise to the 6-cardinal-direction representation 
of Sec.~\ref{sec:pauli}.  The regular polyhedra by no means exhaust 
the possibilities for representations of this sort.  One can use simultaneously 
the vertices from any number of polyhedra.  Moreover, one can produce an 
appropriate set of unit vectors by the following procedure: starting with 
any set of unit vectors in the first octant, reflect the set successively 
through the three Cartesian planes to produce a set of vectors that occupies 
all eight octants.

For more than one qubit, the possibilities for discrete representations
that follow from the continuous representation are even more diverse.
The simplest possibility is to use the same set of pure-state projectors 
for each qubit, but this is not necessary.  One can use a different set 
of projectors for each qubit, or more generally, one can use a ``tree 
structure,'' in which there is a different set of projectors for the $n$th 
qubit  for each projector for the first $n-1$ qubits. There are 
possibilities more general even than those with a tree structure, analogous 
to the ``domino states'' introduced for two qutrits in \cite{Bennett1999a}.  
We do not dwell here on the myriad of discrete representations of this sort, 
merely noting that since the expansion coefficients in such representations 
follow from the expansion coefficients for the continuous representation, 
all such representations lead to the same general lower 
bound~(\ref{eq:bound2}) on the size of the separable neighborhood surrounding 
the maximally mixed state.  Both the continuous and these discrete 
representations can, however, give better bounds on the separability of 
{\it particular\/} states of the form~(\ref{eq:rhoeps}), a point to which 
we return in Sec.~\ref{sec:epcat}.

\section{Spherical-harmonic expansions}
\label{sec:spherical}

In this section we explore the relationship between product-state 
representations and sphe\-ri\-cal-harmonic expansions.

\subsection{Spherical-harmonic expansions and the Pauli representation}
\label{sec:sphpauli}

We begin by noting that if $\n$ is a unit vector, then
\begin{eqnarray}
  \vec\sigma\cdot\n & = & \sigma_3 \cos\theta + \sigma_1\sin\theta\cos\phi
                          + \sigma_2\sin\theta\sin\phi \nonumber\\
  & = & \sqrt{4\pi\over3}\Bigl(\sigma_3{Y_1^0}^\ast 
        + \sigma_1{1\over\sqrt2}(-{Y_1^{+1}}^\ast+{Y_1^{-1}}^\ast)
        - i\sigma_2{1\over\sqrt2}({Y_1^{+1}}^\ast+{Y_1^{-1}}^\ast) \Bigr) \\
  & = & \sigma_1^0{Y_1^0}^\ast+
        \sigma_1^{+1}{Y_1^{+1}}^\ast+\sigma_1^{-1}{Y_1^{-1}}^\ast \;, \nonumber
\end{eqnarray}
where 
\begin{equation}
  Y_0^0={1\over\sqrt{4\pi}}\;,\;\;\;
  Y_1^0=\sqrt{3\over4\pi}\cos\theta\;,\;\;\;
  Y_1^{\pm1}=\mp\sqrt{3\over8\pi}\sin\theta\,e^{\pm i\phi}
\end{equation}
are spherical harmonics and, in the last line, we define
\begin{equation}
  \sigma_0^0\equiv\sqrt{4\pi}\,1\;,\;\;\;
  \sigma_1^0\equiv\sqrt{4\pi\over3}\sigma_3\;,\;\;\;
  \sigma_1^{\pm1}\equiv\mp\sqrt{2\pi\over3}(\sigma_1\pm i\sigma_2)\;.
\end{equation}
We can then write any pure-state projector on the Bloch sphere in the form
\begin{equation}
  P_{\n} = \proj{\n} = {1\over2}(1+\vec\sigma\cdot\n) 
  = {1\over2}\, \sum_{l=0}^1\sum_{m=-l}^l \sigma_l^mY_l^{m*} \;.
\end{equation}

Any $N$-qubit density operator can be written in the form 
\begin{equation}
  \label{eq:rhow}
  \rho = \int \dOn w(\tilde n) P(\tilde n) \;,
\end{equation}
where the expansion coefficients $w(\tilde n)\equiv w(\n_1,\ldots,\n_N)$ are 
not necessarily nonnegative [the expansion coefficients $w(\tilde n)$ need
not be those of the canonical representation~(\ref{eq:wn})].  A density 
operator $\rho$ is separable if and only if it can be written in the form 
(\ref{eq:rhow}) for some nonnegative probability density $w(\tilde n)$. Now 
expand the product projector~(\ref{eq:prodproj}) in terms of spherical
harmonics,
\begin{equation}
  P(\tilde n) = {1\over2^N}\, \sum_{\{l_j\le1,m_j\}}
                Y_{l_1}^{m_1*}\cdots Y_{l_N}^{m_N*}
                \sigma_{l_1}^{m_1}\otimes\cdots\otimes\sigma_{l_N}^{m_N} \;,
\label{eq:projexpand}
\end{equation}
where 
\begin{equation}
  \sum_{\{l_j\le1,m_j\}} \equiv \;
  \sum_{l_1=0}^1\sum_{m_1=-l_1}^{l_1} \cdots 
  \sum_{l_N=0}^1\sum_{m_N=-l_N}^{l_N}
\end{equation}
denotes a sum that contains only scalar and vector terms ($l=0,1$), and also
expand the expansion coefficients,
\begin{equation}
  w(\tilde n) =
  \sum_{\{l_j,m_j\}}
  a_{l_1\ldots l_N}^{m_1\ldots m_N}Y_{l_1}^{m_1}\cdots Y_{l_N}^{m_N} \;,
\label{eq:wexpand}
\end{equation}
where 
\begin{equation}
  \sum_{\{l_j,m_j\}} \equiv \;
  \sum_{l_1=0}^\infty\sum_{m_1=-l_1}^{l_1} \cdots 
  \sum_{l_N=0}^\infty\sum_{m_N=-l_N}^{l_N} 
\end{equation}
denotes an unrestricted sum over spherical harmonics.  Inserting 
Eqs.~(\ref{eq:projexpand}) and (\ref{eq:wexpand}) into Eq.~(\ref{eq:rhow})
and using the orthonormality of the spherical harmonics yields the (unique)
Pauli-operator representation of $\rho$:
\begin{equation}
  \rho= {1\over2^N} \sum_{\{l_j\le1,m_j\}}
  a_{l_1\ldots l_N}^{m_1\ldots m_N}
  \sigma_{l_1}^{m_1}\otimes\cdots\otimes\sigma_{l_N}^{m_N} \;.
  \label{eq:rho2}
\end{equation}

We see that $\rho$ depends on and determines only the scalar and vector parts 
of $a_{l_1\ldots l_N}^{m_1\ldots m_N}$; the higher-order-spherical-harmonic 
(HOSH) content of $w(\tilde n)$ corresponds to the freedom in constructing
pure-product-state representations of $\rho$.  It is straightforward to 
generate all pure-product-state representations: given the Pauli-operator 
representation~(\ref{eq:rho2}) of $\rho$, any $w(\tilde n)$ is given by 
Eq.~(\ref{eq:wexpand}), where the coefficients 
$a_{l_1\ldots l_N}^{m_1\ldots m_N}$ are arbitrary except for 
$l_1,\ldots,l_N\le1$. 

The canonical continuous representation $\wc(\tilde n)$ of Eq.~(\ref{eq:wn})
is the unique pure-product-state representation given by the $l=0$ and
$l=1$ components.  We can see this directly in the following way.  Write
the Pauli representation of $\rho$ in the form
\begin{equation}
  \rho= {1\over2^N }
  \sum_{\alphaseq}\cseq\sigmaseq \;,
  \label{eq:pauli2}
\end{equation}
where
\begin{equation}
  \label{eq:paulic}
  \cseq=\tr(\rho\sigmaseq)=\langle\sigmaseq\rangle \;.
\end{equation}
Here $\sigma_0=1$, and the sums over Greek indices run from 0 to 3.  Now we 
have that
\begin{eqnarray}
  \wc(\tilde n)&=&
    {1\over(4\pi)^N}
    \Bigl\langle 
    (1+3\vec\sigma\cdot\n_1)\otimes\cdots\otimes(1+3\vec\sigma\cdot\n_N)
    \Bigr\rangle \nonumber \\
  \label{eq:probability}
  &=&
     \left({3\over4\pi}\right)^N
     \sum_{\alphaseq}
     \Bigl\langle
     (n_1)_{\alpha_1}\sigma_{\alpha_1}
     \otimes\cdots\otimes
     (n_N)_{\alpha_N}\sigma_{\alpha_N}
     \Bigr\rangle  \\ 
  &=&
     \left({3\over4\pi}\right)^N
     \sum_{\alphaseq}
     \cseq\nseq \;, \nonumber
\end{eqnarray}
where $(n_j)_{\alpha_j}\equiv1/3$ if $\alpha_j=0$ and $(n_j)_{\alpha_j}$ is
a Cartesian component of $\n_j$ if $\alpha_j=1,2$, or 3.  Thus any 
pure-product-state representation of a density operator $\rho$ can be
obtained by adding HOSH to the canonical continuous representation
$\wc(\tilde n)$.  The discrete representations discussed in 
Sec.~\ref{sec:discrete} are examples---though by no means the only 
examples---of representations with HOSH content.

The continuous canonical representation can be generated directly from 
the Pauli representation of $\rho$ by making the substitutions
\begin{equation}
  \label{eq:rule1}
  2^N\rho\longrightarrow \wc(\tilde n) \;,
\end{equation}
\begin{equation}
  \label{eq:rule2}
  1\longrightarrow{1\over4\pi}\;,\;\;\;
  \sigma_j\longrightarrow{3\over4\pi}n_j \;.
\end{equation}
These rules are equivalent to $\sigma_l^m\longrightarrow Y_l^m$ and to
$\sigma_\alpha\longrightarrow(3/4\pi)n_\alpha$.

\subsection{The $\epsilon$-cat state}
\label{sec:epcat}

As an example, consider the following $N$-qubit generalization of the Werner 
\cite{Werner1989a} state,
\begin{equation}
  \rho_\epsilon = {1-\epsilon\over2^N}1 + \epsilon\proj{\cat} \;,
  \label{eq:epcat}
\end{equation}
where 
\begin{equation}
  \label{eq:cat}
  |\cat\rangle \equiv 
  {1\over\sqrt2}\Bigl(\ket{0\ldots0}+\ket{1\ldots1}\Bigr) 
\end{equation}
is the $N$-qubit ``cat state.''  We call the mixed state~(\ref{eq:epcat}) the
{\it $\epsilon$-cat state}.

Now let $\ket0$and $\ket1$ be the eigenstates of $\sigma_3$, with $\ket0$
at the top of the Bloch sphere and $\ket1$ at the bottom, so that 
$\sigma_1=\outer 0 1 +\outer 1 0$, $\sigma_2=-i\outer 0 1 +i\outer 1 0$,
and $\sigma_3=\proj 0 -\proj 1$.  It follows that the Pauli representation
of the $\epsilon$-cat state is
\begin{eqnarray}
  \rho_\epsilon &=& 
     {1\over2^N}\biggl(
        (1-\epsilon)1 \nonumber \\
        &\mbox{}&\hphantom{{1\over2^N}}
        +{\epsilon\over2}(1+\sigma_3)\otimes\cdots\otimes(1+\sigma_3)
        +{\epsilon\over2}(1-\sigma_3)\otimes\cdots\otimes(1-\sigma_3)
        \label{eq:pauliepcat} \\
        &\mbox{}&\hphantom{{1\over2^N}}
        +{\epsilon\over2}
        (\sigma_1+i\sigma_2)\otimes\cdots\otimes(\sigma_1+i\sigma_2)
        +{\epsilon\over2}
        (\sigma_1-i\sigma_2)\otimes\cdots\otimes(\sigma_1-i\sigma_2)
     \biggr) \;. \nonumber
\end{eqnarray}
Applying the rules~(\ref{eq:rule1}) and (\ref{eq:rule2}) yields the 
continuous canonical representation of $\rho_\epsilon$:
\begin{eqnarray}
  \wc_\epsilon(\tilde n) &=& 
    {1\over(4\pi)^N} \biggl(
       (1-\epsilon) \nonumber \\ 
       &\mbox{}&\hphantom{{1\over(4\pi)^N}}
       +{\epsilon\over2}(1+3\cos\theta_1)\cdots(1+3\cos\theta_N) 
       +{\epsilon\over2}(1-3\cos\theta_1)\cdots(1-3\cos\theta_N) 
       \nonumber \\
       &\mbox{}&\hphantom{{1\over(4\pi)^N}}
       +\epsilon\,3^N
       \sin\theta_1\cdots\sin\theta_N\cos(\phi_1+\ldots+\phi_N)
    \biggr) \;. 
    \label{eq:wcat} 
\end{eqnarray}

The minimum value of $\wc_\epsilon(\tilde n)$ occurs either (i)~when one qubit
is evaluated at the north (south) pole, and the other $N-1$ qubits are evaluated 
at the south (north) pole, or (ii)~when all the qubits are evaluated on the 
equator with $\cos(\phi_1+\ldots+\phi_N)=-1$.  For $N=2$ and $N=4$, both the 
poles and the equator give the same minimum value, which shows that 
the $\epsilon$-cat state is separable for $\epsilon\le1/9$ for $N=2$ and for 
$\epsilon\le1/81$ for $N=4$.  For $N=3$ and $N=5$, the equator gives the 
minimum value, which shows that the $\epsilon$-cat state is separable for 
$\epsilon\le1/27$ for $N=3$ \cite{Braunstein1999a} and for 
$\epsilon\le1/243$ for $N=5$.  For $N\ge6$, the poles give the minimum 
value, which shows that $\rho_\epsilon$ is separable for
\begin{equation}
  \epsilon\le{1\over1\pm 2^N+2^{2N-2}} \;,
\label{eq:sepNcat}
\end{equation}
the upper (lower) sign applying when $N$ is even (odd).  Notice that except
for $N=2$, all these bounds are better than the general-purpose bound
of Eq.~(\ref{eq:bound2}).

These bounds on the separability of the $\epsilon$-cat state can be improved
by using tetrahedral ensembles of the sort discussed in Sec.~\ref{sec:discrete}.
The tetrahedral ensembles are chosen to have a tree structure such that the
vertices avoid the places where $\wc_\epsilon(\tilde n)$ has its smallest
values.  We do not discuss these tetrahedral bounds here, preferring instead
to consider in detail the separability of the $\epsilon$-GHZ state.

\section{The $\epsilon$-GHZ state}
\label{sec:GHZ}

In this section we consider the 3-qubit version of the $\epsilon$-cat
state introduced in Sec.~\ref{sec:epcat}.  In the case of 3 qubits, the 
cat state of Eq.~(\ref{eq:cat}) is called the Greenberger-Horne-Zeilinger 
(GHZ) state \cite{Greenberger1990},
\begin{equation}
  |\psi_{\rm GHZ}\rangle = {1\over\sqrt2}\Bigl(\ket{000}+\ket{111}\Bigr) \;;
\end{equation} 
we call the mixture $\rho_\epsilon$ of Eq.~(\ref{eq:epcat}) the 
{\it $\epsilon$-GHZ state}.  We show in this section that {\it the 
$\epsilon$-GHZ state is separable if and only if $\epsilon\le1/5$}.  
The class of $\epsilon$-GHZ states has been considered previously in 
\cite{Murao1998}.

\subsection{Separability of the Werner state}
\label{sec:werner}

Before proceeding to the proof of the separability of the $\epsilon$-GHZ
state, we review a similar, but simpler proof for the Werner state 
\cite{Werner1989a}, the 2-qubit version of the $\epsilon$-cat state:
\begin{eqnarray}
  \rho_W &=& {1-\epsilon\over4}1\otimes1+
  {\epsilon\over2}
  \Bigl(|00\rangle+|11\rangle\Bigr)\Bigl(\langle00|+\langle11|\Bigr) 
  \nonumber \\
  &=& {1\over4}\Bigl(1\otimes1+\epsilon\,
     (\sigma_3\otimes\sigma_3+\sigma_1\otimes\sigma_1-\sigma_2\otimes\sigma_2)
     \Bigr)\;.
  \label{eq:rhowerner}
\end{eqnarray}
The Werner state is separable if and only if $\epsilon\le1/3$ 
\cite{Bennett1996a,RHorodecki1996a}, a result we now show in the following
way.  

A density operator $\rho$ for 2 qubits, $A$ and $B$, is separable if and only 
if it can be written as
\begin{equation}
  \rho=\int\dOnAnB\wnAnB\PnAnB
\;,
\end{equation}
with $\wnAnB$ being a nonnegative probability distribution.  If a state is 
separable, the coefficients~(\ref{eq:paulic}) in the Pauli 
representation~(\ref{eq:pauli2}) of $\rho$ become classical expectations
over the distribution $\wnAnB$, expectations that we denote by $E[\cdot]$:
\begin{equation}
  c_{\alpha\beta}=
  \tr(\rho\,\sigma_\alpha\otimes\sigma_\beta)
  =\int\dOnAnB(n_A)_\alpha(n_B)_\beta\,\wnAnB\equiv
  E[(n_A)_\alpha(n_B)_\beta] \;.
\end{equation}

For the Werner state, the coefficients $c_{\alpha\beta}$ can be read off 
the Pauli representation in Eq.~(\ref{eq:rhowerner}):
\begin{eqnarray}
   &\mbox{}&E[(n_A)_j]=E[(n_B)_j]=0 \;, \\
   &\mbox{}&E[(n_A)_j(n_B)_k]=0 \;,\;\;\mbox{for $j\ne k$,} \\
   \label{eq:diag}
   &\mbox{}&E[(n_A)_1(n_B)_1]=-E[(n_A)_2(n_B)_2]=E[(n_A)_3(n_B)_3]=\epsilon \;.
\end{eqnarray}
If $\wnAnB$ is a probability distribution, we have that
\begin{equation}
  \epsilon=\Bigl|E[(n_A)_j(n_B)_j]\Bigr|\le
  {1\over2}E[(n_A)_j^2+(n_B)_j^2] \;.
\end{equation}
Adding over the three values of $j$ gives
\begin{equation}
  3\epsilon \le {1\over2}
  E[(n_A)_1^2+(n_A)_2^2+(n_A)_3^2+(n_B)_1^2+(n_B)_2^2+(n_B)_3^2]=1 \;,
\end{equation}
which shows that if $\epsilon>1/3$, the Werner state is nonseparable.

We still have the task of designing a product ensemble that gives the Werner
state for $\epsilon\le1/3$. Defining three density operators
\begin{eqnarray}
  \rho_1 &\equiv&
  {1\over2}\bigl(
     P_{\vec e_3}\otimes P_{\vec e_3}+P_{-\vec e_3}\otimes P_{-\vec e_3}
     \bigr)
  ={1\over4}\bigl(
     1\otimes1+\sigma_3\otimes\sigma_3
     \bigr) \;, \\
  \rho_2 &\equiv&
  {1\over2}\bigl(
     P_{\vec e_1}\otimes P_{\vec e_1}+P_{-\vec e_1}\otimes P_{-\vec e_1}
     \bigr)
  ={1\over4}
     \bigl(1\otimes1+\sigma_1\otimes\sigma_1
     \bigr) \;, \\
  \rho_3 &\equiv& 
     {1\over2}\bigl(
     P_{\vec e_2}\otimes P_{-\vec e_2}+P_{-\vec e_2}\otimes P_{\vec e_2}
     \bigr)
  ={1\over4}\bigl(
     1\otimes1-\sigma_2\otimes\sigma_2
     \bigr) \;,
\end{eqnarray}
we see that each takes care of one of the correlations in Eq.~(\ref{eq:diag}).  
An equal mixture gives the $\epsilon=1/3$ Werner state:
\begin{equation}
  {1\over3}(\rho_1+\rho_2+\rho_3)={1\over4}\left(
  1\otimes1+{1\over3}\bigl(
  \sigma_3\otimes\sigma_3+\sigma_1\otimes\sigma_1-\sigma_2\otimes\sigma_2
  \bigl)\right) \;.
\end{equation}
Smaller values of $\epsilon$ can be handled by adding on a product ensemble
for the maximally mixed state.

\subsection{Separability of the $\epsilon$-GHZ state}
\label{sec:epGHZ}

Consider now the $\epsilon$-GHZ state
\begin{eqnarray}
  \rhoGHZ &=& {1-\epsilon\over8}1\otimes1\otimes1+
  {\epsilon\over2}
  \Bigl(|000\rangle+|111\rangle\Bigr)\Bigl(\langle000|+\langle111|\Bigr) 
  \nonumber \\
  \label{eq:rhoepGHZ}
  &=& {1\over8}\biggl(1\otimes1\otimes1+
       \epsilon\Bigl(
       1\otimes\sigma_3\otimes\sigma_3
       +\sigma_3\otimes1\otimes\sigma_3
       +\sigma_3\otimes\sigma_3\otimes1  \\
  &\mbox{}&\hphantom{{1\over8}\Bigl(1}
       +\sigma_1\otimes\sigma_1\otimes\sigma_1
       -\sigma_1\otimes\sigma_2\otimes\sigma_2
       -\sigma_2\otimes\sigma_1\otimes\sigma_2
       -\sigma_2\otimes\sigma_2\otimes\sigma_1
       \Bigr)\biggr) \;. \nonumber
\end{eqnarray}

A density operator $\rho$ for 3 qubits, $A$, $B$, and $C$, is separable 
if and only if it can be written as
\begin{equation}
\rho=\int\dO\wn\Pn
\;,
\end{equation}
with $\wn$ being a nonnegative probability distribution.  If a state is 
separable, the coefficients~(\ref{eq:paulic}) in the Pauli 
representation~(\ref{eq:pauli2}) of $\rho$ become classical expectations
over the distribution $\wn$: 
\begin{equation}
  c_{\alpha\beta\gamma} =
  \tr(\rho\,\sigma_\alpha\otimes\sigma_\beta\otimes\sigma_\gamma) 
  = E[(n_A)_\alpha(n_B)_\beta(n_C)_\gamma] \;. 
\end{equation}

For the $\epsilon$-GHZ state, the coefficients $c_{\alpha\beta\gamma}$ can 
be read off the Pauli representation in Eq.~(\ref{eq:rhoepGHZ}):
\begin{eqnarray}
  &\mbox{}&E[(n_A)_j]=E[(n_B)_j]=E[(n_C)_j]=0 \;, \\
  \label{eq:corr3}
  &\mbox{}&E[(n_A)_j(n_B)_k]=E[(n_A)_j(n_C)_k]=E[(n_B)_j(n_C)_k] =
    \epsilon\,\delta_{j3}\delta_{k3} \;, \\
  \label{eq:corr12}
  &\mbox{}&E[(n_A)_j(n_B)_k(n_C)_l]=\epsilon\,(
    \delta_{j1}\delta_{k1}\delta_{l1}-\delta_{j1}\delta_{k2}\delta_{l2}
    -\delta_{j2}\delta_{k1}\delta_{l2}-\delta_{j2}\delta_{k2}\delta_{l1}) \;.
\end{eqnarray}
Now define a ``vector''
\begin{equation}
  \vec N\equiv
  \Bigl((n_B)_1(n_C)_1-(n_B)_2(n_C)_2\Bigr)\vec e_1
  -\Bigl((n_B)_1(n_C)_2+(n_B)_2(n_C)_1\Bigr)\vec e_2
  +(n_B)_3\vec e_3 \;,
\end{equation}
whose magnitude satisfies
\begin{eqnarray}
  \vec N\cdot\vec N &=&
  \Bigl((n_B)_1(n_C)_1-(n_B)_2(n_C)_2\Bigr)^2+
  \Bigl((n_B)_1(n_C)_2+(n_B)_2(n_C)_1\Bigr)^2+(n_B)_3^2 \nonumber \\
  &=& \Bigl((n_B)_1^2+(n_B)_2^2\Bigr)\Bigl((n_C)_1^2+(n_C)_2^2\Bigr)
  +(n_B)_3^2  \\
  &\le& \n_B\cdot\n_B=1 \;. \nonumber
\end{eqnarray}
If $\wn$ is a probability distribution, we have that
\begin{equation}
  5\epsilon=
  \Bigl|E\Bigl[\n_A\cdot\vec N\,\Bigr]\Bigr| 
  \le E\Bigl[\Bigl|\n_A\cdot\vec N\Bigr|\Bigr]
  \le {1\over2}E\Bigl[\n_A\cdot\n_A+\vec N\cdot\vec N\,\Bigr] \le 1 \;,
\end{equation}
which shows that if $\epsilon>1/5$, the $\epsilon$-GHZ state is nonseparable.

To finish the proof, we need to construct a product ensemble for the
$\epsilon$-GHZ state when $\epsilon\le1/5$.  We begin by defining five
density operators
\begin{eqnarray}
  \rho_1 &\equiv& {1\over2}\Bigl(
     \Pe{}{3}{}{3}{}{3}+\Pe{-}{3}{-}{3}{-}{3} \Bigr) \nonumber \\
  &=& {1\over8}\Bigl( 
     1\otimes1\otimes1+1\otimes\sigma_3\otimes\sigma_3
     +\sigma_3\otimes1\otimes\sigma_3+\sigma_3\otimes\sigma_3\otimes1
     \Bigr) \;, \\
  \rho_2 &\equiv& {1\over4}\Bigl(
     \Pe{}{1}{}{1}{}{1}+ \Pe{}{1}{-}{1}{-}{1}+ 
     \Pe{-}{1}{}{1}{-}{1}+ \Pe{-}{1}{-}{1}{}{1} \Bigr) \nonumber \\
  &=&{1\over8}\bigl(
     1\otimes1\otimes1+\sigma_1\otimes\sigma_1\otimes\sigma_1 \Bigr) \;, \\
  \rho_3 &\equiv& {1\over4}\Bigl(
     \Pe{}{1}{}{2}{-}{2}+ \Pe{}{1}{-}{2}{}{2}+ 
     \Pe{-}{1}{}{2}{}{2}+ \Pe{-}{1}{-}{2}{-}{2} \Bigr) \nonumber \\
  &=& {1\over8}\bigl(
     1\otimes1\otimes1-\sigma_1\otimes\sigma_2\otimes\sigma_2 \Bigr) \;, \\
  \rho_4 &\equiv& {1\over4}\Bigl(
     \Pe{}{2}{}{1}{-}{2}+ \Pe{-}{2}{}{1}{}{2}+
     \Pe{}{2}{-}{1}{}{2}+ \Pe{-}{2}{-}{1}{-}{2} \Bigr) \nonumber \\
  &=& {1\over8}\Bigl(
     1\otimes1\otimes1-\sigma_2\otimes\sigma_1\otimes\sigma_2 \Bigr) \;, \\
  \rho_5 &\equiv& {1\over4}\Bigl(
     \Pe{}{2}{-}{2}{}{1}+ \Pe{-}{2}{}{2}{}{1}+
     \Pe{}{2}{}{2}{-}{1}+ \Pe{-}{2}{-}{2}{-}{1} \Bigl) \nonumber \\
  &=& {1\over8}\Bigl(
     1\otimes1\otimes1- \sigma_2\otimes\sigma_2\otimes\sigma_1 \Bigr) \;,
\end{eqnarray}
each of which takes care of one of the nonzero correlations in 
Eqs.~(\ref{eq:corr3}) and (\ref{eq:corr12}).  An equal mixture gives the 
($\epsilon=1/5$)-GHZ state:
\begin{eqnarray}
  &\mbox{}&{1\over5}(\rho_1+\rho_2+\rho_3+\rho_4+\rho_5) \nonumber \\
  &\mbox{}&\hphantom{{1\over5}(\rho}
     ={1\over8}\biggl(1\otimes1\otimes1+
      {1\over5}\Bigl(1\otimes\sigma_3\otimes\sigma_3
      +\sigma_3\otimes1\otimes\sigma_3+\sigma_3\otimes\sigma_3\otimes1 \\
  &\mbox{}&\hphantom{{1\over5}(\rho={1\over8}\biggl(1}
      +\sigma_1\otimes\sigma_1\otimes\sigma_1 
      -\sigma_1\otimes\sigma_2\otimes\sigma_2
      -\sigma_2\otimes\sigma_1\otimes\sigma_2
      -\sigma_2\otimes\sigma_2\otimes\sigma_1 \Bigr)\biggr) \;, \nonumber
\end{eqnarray}
thus completing the proof.

For both the Werner state and the $\epsilon$-GHZ states, we are able to
construct an optimal ensemble based on the 6 cardinal directions, but
notice that this ensemble is not the canonical representation of 
Sec.~\ref{sec:pauli}.

\section{Conclusion}
\label{sec:discuss}

In this paper we introduce a canonical procedure for constructing 
representations of density operators in terms of complete or overcomplete
sets of pure product projectors.  The procedure is useful for investigating
the separability of density operators: if the expansion coefficients in 
such a representation are nonnegative, they can be interpreted as a 
probability distribution, with the result that the representation then 
provides an explicit product ensemble for the density operator.  Generally, 
however, the canonical representations have negative values even when the 
density operator is separable.  

The canonical continuous representation is the unique representation that
contains only scalar and vector spherical-harmonic terms, and we outline
how to obtain all possible pure-product-state representations from the 
canonical continuous representation by adding arbitrary 
higher-order-spherical-harmonic content.  Searching the space of 
spherical-harmonic expansions for nonnegative representations would be 
quite difficult, but we provide an optimal ensemble in the particular case 
of the $\epsilon$-GHZ state, thereby identifying the separability boundary 
for this class of states.

\section*{Acknowledgments}

Thanks to G.~J. Milburn and R.~Pranaw for discussions of the method used
to prove the separability of the $\epsilon$-GHZ state.  This work was 
supported in part by the US Office of Naval Research 
(Grant No.~N00014-93-0-0116) and by the UK Engineering and Physical Sciences 
Research Council.

\section*{Note added}

The separability bound for the $\epsilon$-GHZ state derived
in Sec.~\ref{sec:GHZ} has also been found by D\"ur, Cirac, and
Tarrach \cite{Duer1999} as part of a more general result. They prove
that the $N$-qubit $\epsilon$-cat state (\ref{eq:epcat}) is separable
if and only if $\epsilon<1/(1+2^{N-1})$, which supersedes our
Eq.~(\ref{eq:sepNcat}). 


\begin{thebibliography}{1}

\bibitem{Peres1996a}
A.~Peres, Phys.\ Rev.\ Lett.\ {\bf 77},  1413  (1996).

\bibitem{Horodecki1996a}
M.~Horodecki, P.~Horodecki, and R.~Horodecki, Phys.\ Lett.~A {\bf 223},  1
  (1996).

\bibitem{Zyczkowski1998}
K.~\.{Z}yczkowski, P.~Horodecki, A.~Sanpera, and M.~Lewenstein, Phys.\ Rev.\ A
  {\bf 58},  883  (1998).

\bibitem{Braunstein1999a}
S.~L. Braunstein, C.~M. Caves, R.~Jozsa, N.~Linden, S.~Popescu, and 
R.~Schack, submitted to Phys. Rev. Lett. {\tt (quant-ph/9811018)}.

\bibitem{Cory1997}
D.~G. Cory, A.~F. Fahmy, and T.~F. Havel, Proc.\ Nat.\ Acad.\ Sci.\ USA 
  {\bf 94},  1634  (1997).

\bibitem{Gershenfeld1997}
N.~A. Gershenfeld and I.~L. Chuang, Science {\bf 275},  350  (1997).

\bibitem{Cory1998}
D.~G. Cory, M.~D. Price, and T.~F. Havel, Physica~D {\bf 120}, 82 (1998).

\bibitem{Greenberger1990}
D.~M. Greenberger, M.~A. Horne, A.~Shimony, and A.~Zeilinger,
Am.\  J.\ Phys. {\bf 58}, 1131 (1990).

\bibitem{Caves1999a}
C.~M. Caves, J.\ Superconductivity, to be published {\tt (quant-ph/9811082)}.

\bibitem{Bennett1999a}
C.~H. Bennett, D.~P. DiVincenzo, C.~A. Fuchs, T.~Mor, E.~Rains, P.~W. Shor,
J.~A. Smolin, and W.~K. Wootters, Phys.\ Rev.~A {\bf 59}, 1070 (1999).

\bibitem{Werner1989a}
R.~F. Werner, Phys.\ Rev.~A {\bf 40}, 4277 (1989).

\bibitem{Murao1998}
M. Murao, M. B. Plenio, S. Popescu, V. Vedral, P. L. Knight, 
Phys.\ Rev.\ A {\bf 57},  R4075  (1998).

\bibitem{Bennett1996a}
C.~H. Bennett, G.~Brassard, S.~Popescu, B.~Schumacher, J.~A. Smolin, and 
W.~K. Wootters, Phys.\ Rev.\ Lett.\ {\bf 76}, 722 (1996).

\bibitem{RHorodecki1996a}
R.~Horodecki and M.~Horodecki, Phys.\ Rev.~A {\bf 54}, 1838 (1996).

\bibitem{Duer1999}
W. D\"ur, J. I. Cirac, and R. Tarrach, {\tt quant-ph/9903018}.

\end{thebibliography}

\end{document}